\documentclass[
]{ceurart}

\usepackage{listings}
\usepackage{tabularx}
\usepackage{graphicx}
\usepackage{subcaption}
\usepackage{enumitem}
\usepackage{todonotes}

\definecolor{link}{RGB}{0, 123, 255}

\begin{document}

\copyrightyear{2026}
\copyrightclause{Copyright for this paper by its authors. Use permitted under Creative Commons License Attribution 4.0 International (CC BY 4.0).}

 \conference{ISWC 2026 Companion Volume, October 25--29, 2026, Bari, Italy}

\title{SciKGDash: The Human-in-the-Loop Curation Dashboard for the Open Research Knowledge Graph}


\author[1]{Lena John}[%
orcid=0009-0007-2097-9761,
email=lena.john@tib.eu,
]
\address[1]{
TIB - Leibniz Information Centre for Science and Technology, Hannover, Germany}

\author[1]{Sören Auer}[%
orcid=0000-0002-0698-2864,
email=auer@tib.eu,
]

\author[1]{Oliver Karras}[%
orcid=0000-0001-5336-6899,
email=oliver.karras@tib.eu,
]

\begin{abstract}
\textit{\textbf{[Background.]}} Research knowledge graphs (RKGs) represent complex scientific knowledge in a machine-actionable format, but their downstream utility depends heavily on their quality. In crowdsourced environments, such as the Open Research Knowledge Graph (ORKG), non-restrictive modeling freedom leads to structural dispersion and anomalies, such as semantic inconsistencies, duplicate predicates, and incomplete metadata. As the ORKG is mainly designed for content creation, it lacks dedicated curation workflows, burdening the ORKG Curation \& Community Building (C\&CB) team with inefficient manual detection and resolution processes.
\textbf{[Objective.]} Our goal is to bridge the gap between non-restrictive crowdsourced modeling and semantic structural rigor by providing \textit{SciKGDash}. Although it is a dedicated platform for human-in-the-loop (HITL) curation to systematically scale RKG quality assurance for the ORKG, it also serves as a reference framework for the broader RKG landscape.
\textbf{[Method.]} Building on a Minimum Viable Product (MVP), we advanced \textit{SciKGDash} into a production-ready system with a decoupled, containerized architecture spanning three functional tiers: Strategic (macro-level monitoring), Analytical (proactive usage analytics), and Operational (HITL curation workflows). The dashboard is publicly available as an open-source project and evolves through continuous action research with the C\&CB team to iteratively integrate new requirements.
\textbf{[Results.]} \textit{SciKGDash} provides comprehensive RKG quality assurance with the following core capabilities: (1) Macro-level monitoring to detect systemic compliance gaps and anomalies; (2) Proactive usage analytics to transform anonymized usage data into actionable design insights; and (3) HITL curation workflows to identify and resolve compliance gaps and anomalies at scale.
\textbf{[Conclusions.]} \textit{SciKGDash} establishes a scalable foundation for HITL curation for the ORKG. It demonstrates how targeted RKG quality assurance can effectively bridge the gap between non-restrictive crowdsourced modeling and semantic structural rigor essential for reliable, machine-actionable scientific knowledge.
\end{abstract}


\maketitle
\section{Introduction}
Research knowledge graphs (RKGs) as specialized form of Knowledge Graphs (KGs) represent complex scientific knowledge in a machine-actionable format~\cite{Zloch.2025,Stocker.2022, Karras.2023, Auer.2025}. However, their downstream utility, e.g., for semantic search or automated synthesis, depends heavily on their quality~\cite{Chen.2019}. Therefore, ensuring high-quality RKGs is crucial for enabling their effective downstream use.

A prominent example illustrating this challenge is the \href{https://orkg.org/}{\textit{\textcolor{link}{Open Research Knowledge Graph}}} (ORKG)~\cite{Auer.2025}, a crowdsourced RKG for creating, organizing, and reusing scientific knowledge. The ORKG prioritizes user freedom by enabling non-restrictive modeling to accommodate diverse research fields~\cite{Auer.2025}. This enables researchers to create and extend knowledge, while supporting its reuse for related-work analysis, visualizations, and querying. While this openness lowers barriers to entry, contributions from diverse global users vary in terms of granularity, structural conformance, and completeness. As a result, the structural dispersion introduces widespread anomalies, including semantic inconsistencies, duplicate predicates, incomplete metadata, and incorrectly typed literals~\cite{John.2026}. Curation is therefore essential to maintain knowledge quality and usability as the ORKG grows. However, the ORKG is primarily designed for content creation rather than curation. Consequently, the \textit{ORKG Curation \& Community Building} (C\&CB) team has to rely on inefficient manual detection and resolution processes due to the lack of dedicated curation workflows.

Our goal is to support open, crowdsourced knowledge creation while ensuring the quality of its semantic representation by providing \href{https://scikgdash.orkg.org/}{\textit{\textcolor{link}{SciKGDash}}}. While researchers contribute and assess the scientific content as domain experts, \textit{SciKGDash} supports curators in checking and improving its semantic modeling quality.
\textit{SciKGDash} is a dedicated platform for human-in-the-loop (HITL) curation to systematically scale RKG quality assurance for the ORKG. Although the dashboard is implemented for the ORKG, it serves as a reference framework for HITL curation in the broader RKG landscape. \textit{SciKGDash} addresses RKG quality assurance through a decoupled, containerized architecture spanning three functional tiers: 1) macro-level monitoring to detect systemic compliance gaps and anomalies; 2) proactive usage analytics to transform anonymized usage data into actionable design insights; and 3) HITL curation workflows to identify and resolve compliance gaps and anomalies at scale.

Following an intial Minimum Viable Product (MVP)~\cite{John_SciKGDash.2025}, \textit{SciKGDash} has advanced into a publicly accessible, production-ready \href{https://scikgdash.orkg.org/}{\textit{\textcolor{link}{system}}} that has been in use since June 23, 2026. Its expanded \href{https://scikgdash.orkg.org/team}{\textit{\textcolor{link}{development team}}} conducts continuous action research to iteratively integrate new requirements from the C\&CB team. This paper presents the progressed state of \textit{SciKGDash}, showcasing its scalable framework, advanced features, and HITL curation workflows that substantiate its efficacy for targeted RKG quality assurance.

\section{Related Work}
Several tools have been proposed to assess and monitor the quality of (R)KGs. \href{https://kgheartbeat.di.unisa.it/}{\textit{\textcolor{link}{KGHeartBeat}}}~\cite{Pellegrino.2025} provides a fully automated framework with a publicly available online service that periodically evaluates (R)KGs from the LOD Cloud and DataHub using comprehensive quality metrics. Similarly, \href{https://github.com/dmki-tuwien/metriKG}{\textit{\textcolor{link}{MetriKG}}}~\cite{Günes.2026} computes a wide range of structural metrics for SPARQL endpoints and RDF datasets. ABSTAT-HD~\cite{Principe.2022} complements metric-based approaches by generating semantic profiles that facilitate dataset exploration and help identify potential quality issues. 
Other tools address specific aspects of quality management. The \href{http://shacl-dashboard-app.eba-yqivtahf.eu-north-1.elasticbeanstalk.com/}{\textit{\textcolor{link}{SHACL Dashboard}}}~\cite{Mäkelburg.2026} enables interactive analysis of large SHACL validation reports. 
In crowdsourced contexts, \href{https://www.mediawiki.org/wiki/ORES}{\textit{\textcolor{link}{ORES}}}~\cite{Ye.2021} applies machine learning to detect damaging Wikipedia edits, and \href{https://integraality.toolforge.org/}{\textit{\textcolor{link}{inteGraality}}} helps identify completeness gaps in Wikidata. Beyond data quality, eFAIR-X~\cite{Bakaki.2026} introduces explainable FAIR-based indicators for assessing computational reproducibility using an RKG.

While these works provide mature solutions for monitoring and validation, a gap remains in the operationalization of these insights. Existing approaches treat RKG quality assurance as a passive detection task without active resolution. We close this ``\textit{detection-to-resolution}'' loop. Our HITL strategy combines monitoring with HITL curation to evolve RKG quality assurance from a periodic diagnostic event into a continuous operational process. In this way, we shift the focus from merely detecting anomalies to orchestrating their resolution at scale. In contrast to these specialized tools, \href{https://scikgdash.orkg.org/}{\textit{\textcolor{link}{SciKGDash}}} integrates diverse quality assessments with interactive exploration and continuous HITL curation workflows, enabling curators to detect, identify, and resolve issues within a single platform.

\section{SciKGDash: Development, Architecture, and Features}
Based on the MVP, we re-engineered \textit{SciKGDash} for scalability, data persistence, advanced features, and HITL curation. The entire \href{https://gitlab.com/TIBHannover/orkg/scikgdash}{\textit{\textcolor{link}{project}}} is open-source and is running as a publicly accessible \href{https://scikgdash.orkg.org/}{\textit{\textcolor{link}{system}}}.


\subsection{Development and Deployment}
For agile development and low downtime operations, the entire application stack is containerized using Docker, allowing developers to set up local environments through a unified configuration. Deployment is automated via a CI/CD pipeline with two dedicated environments and different database instances:

\begin{itemize}[noitemsep, leftmargin=12pt, topsep=0pt]
    \item \textbf{Sandbox Environment:} A \href{https://scikgdash.sandbox.orkg.org/}{\textit{\textcolor{link}{staging instance}}} where experimental or new features, analytical metrics, and schema changes are deployed for integration testing and validation before release.
    \item \textbf{Production Environment:} The \href{https://scikgdash.orkg.org/}{\textit{\textcolor{link}{stable, publicly accessible instance}}} is updated only after changes have been validated in the sandbox environment, ensuring reliable and consistent user experience.
\end{itemize}

The development \textit{SciKGDash} is driven by an expanded development team that conducts continuous action research: When the C\&CB team identifies new requirements, the development team evaluates technical feasibility and translates these requirements into new or modified features. In addition, we finalize a comprehensive system documentation framework to support onboarding and system extension by new development team members. Our specification-driven development approach defines key system components and provides the basis for AI-assisted code generation through controlled harnesses that enforce defined security boundaries.

\subsection{System Architecture and Components}
\begin{figure}[htbp]
    \vspace{-0.4cm}
    \centering
    \fbox{\includegraphics[width=\linewidth]{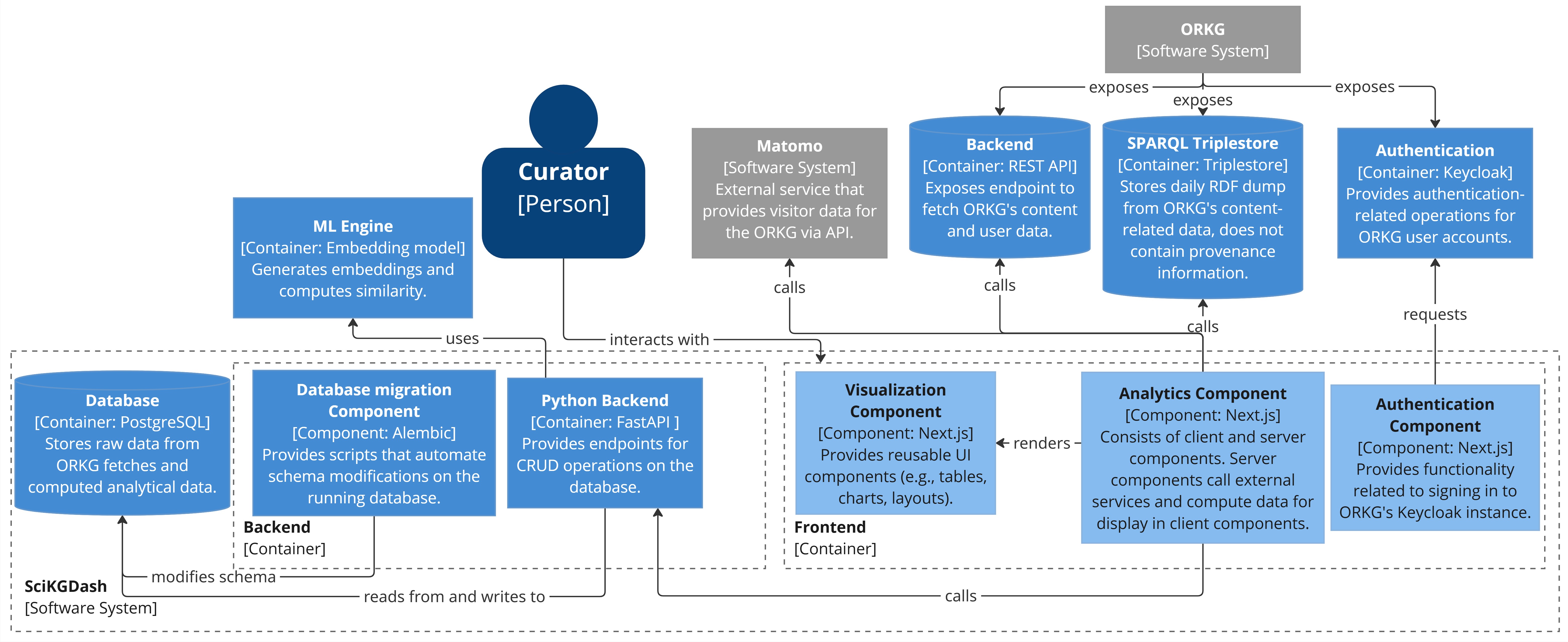}}
    \caption{C4 model -- Container and Component diagram showing the \textit{SciKGDash} system architecture and its interactions with the ORKG and the external Matomo service.}
    \label{fig:c4}
    \vspace{-0.3cm}
\end{figure}

The architecture of \textit{SciKGDash} follows a modular design that separates the frontend, backend, data storage, and external services into independent containers and components (see~\figurename{~\ref{fig:c4}}).\vspace{0.2cm}

\noindent
\textbf{Frontend:} Built with Next.js, the frontend consists of reusable visualization components (e.g., tables, charts, cards), analytics components that retrieve and process data from external services, and an optional authentication component for ORKG login and curator authentication. All content is publicly viewable, while content modification is restricted to authenticated users with curator rights. This architecture enables efficient server-side data processing while providing an interactive user interface.

\noindent
\textbf{Backend:} A lightweight \textit{FastAPI} backend provides REST endpoints for database operations, bulk retrieval of ORKG data, and embedding generation and similarity computation through an external Machine Learning (ML) engine. Database access is handled through an object-relational mapping, while Alembic manages database schema migrations.

\noindent
\textbf{Data Storage and External Services:} The dashboard retrieves data from the ORKG via its REST API and SPARQL endpoint and visitor statistics from the ORKG's Matomo instance via the \href{https://developer.matomo.org/api-reference/api}{\textit{\textcolor{link}{Matomo API}}}. An internal PostgreSQL database persistently stores selected computationally expensive data, allowing these components to serve cached data and refresh it on demand, while the remaining components retrieve current data from the ORKG when loaded. Authentication is delegated to the ORKG's Keycloak instance. Public quality metrics are read-accessible to all users, while write operations, such as executing HITL curation workflows, are restricted to authenticated curators.

\subsection{Features and Curation Workflows}
\begin{figure}
    \centering
    \begin{subfigure}[b]{0.58\textwidth}
        \centering
        \fbox{\includegraphics[width=\linewidth]{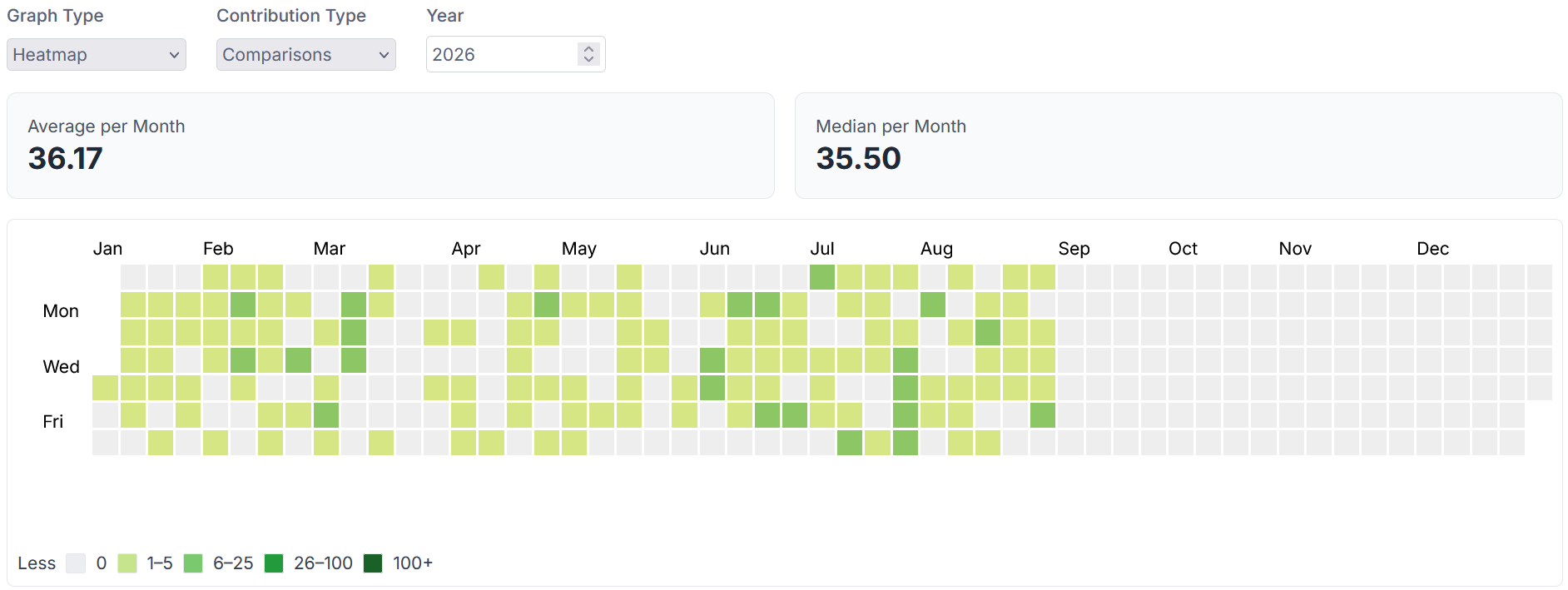}}
        \caption{Strategic tier: Contributions per entity, here comparisons}
        \label{subfig:contributions}
    \end{subfigure}
    \hfill
    \begin{subfigure}[b]{0.4\textwidth}
        \centering
        \fbox{\includegraphics[width=\linewidth]{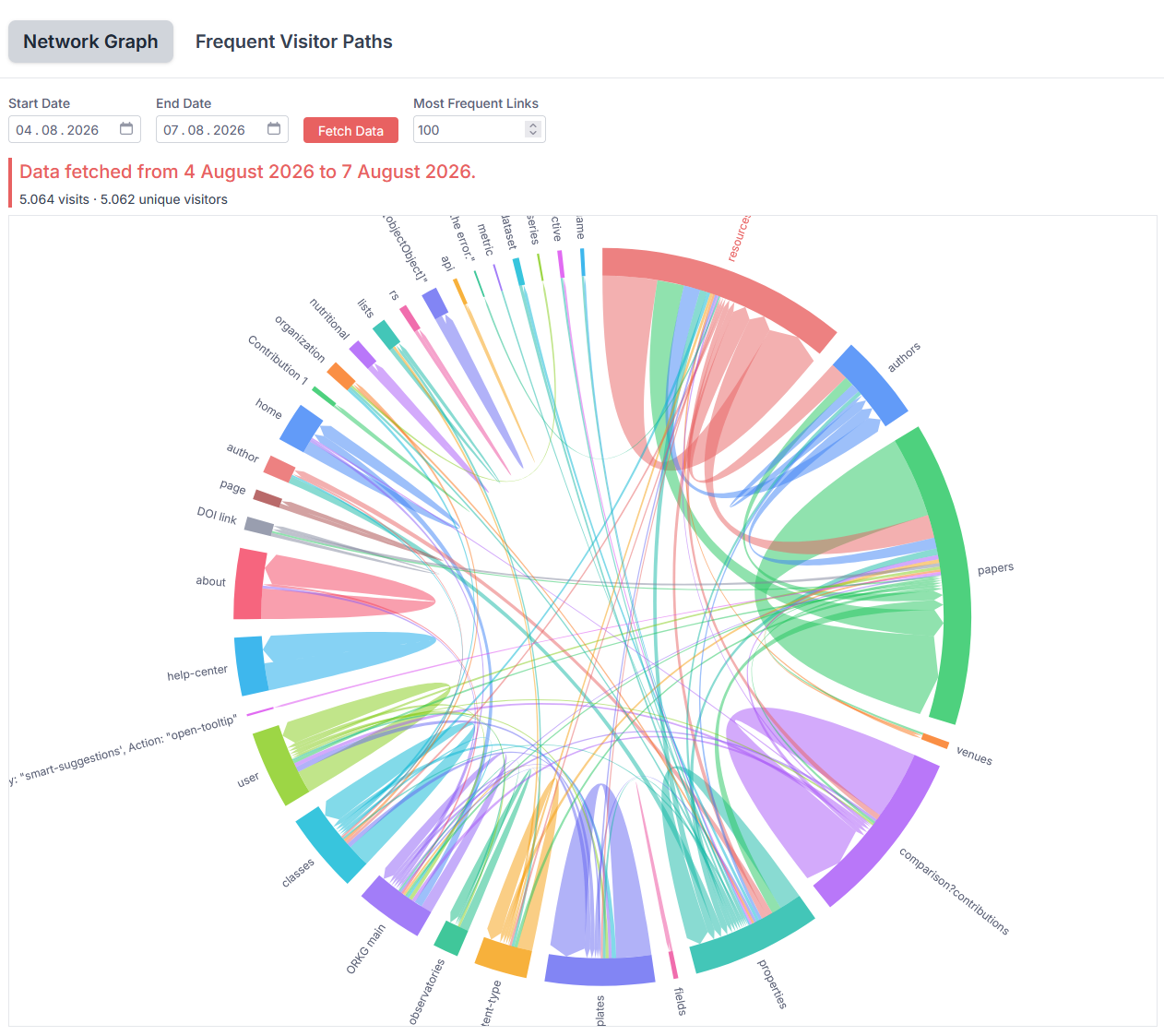}}
        \caption{Analytical tier: Visitor navigation paths}
        \label{subfig:similarity_search}
    \end{subfigure}
    \label{fig:visitors}
    \hfill
    \begin{subfigure}[h]{\textwidth}
        \centering
        \fbox{\includegraphics[width=\linewidth]{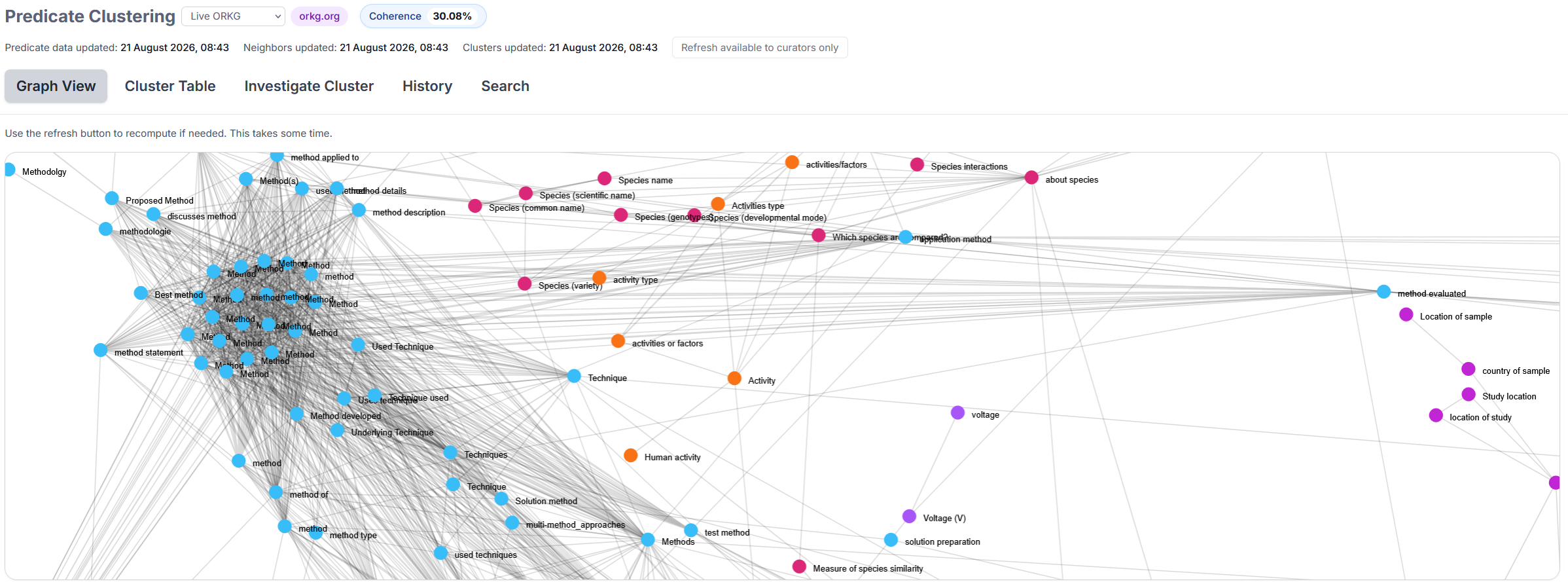}}
        \caption{Operational tier: Duplicate predicate detection, Clustering view}
        \label{subfig:duplicates}
    \end{subfigure}
    \caption{Representative screenshots illustrating the three functional tiers of \textit{SciKGDash}}
    \label{fig:screenshots}
\end{figure}

To systematically scale HITL curation for the ORKG, \textit{SciKGDash} implements three functional tiers: Strategic, Analytical, and Operational~\cite{Few.2005}. These tiers shift curation from ad hoc discovery and manual resolution of individual quality issues to systemic RKG quality assurance at scale. Table~\ref{tab:implemented_features} provides an overview of the supported features, their evolution status from the MVP~\cite{John_SciKGDash.2025}, and the targeted tier. Figure \ref{fig:screenshots} shows three representative screenshots for each tier in \textit{SciKGDash}.

\begin{table}[htbp]
\vspace{-0.3cm}
\centering
\caption{Supported features with links to SciKGDash, categorized by evolution status and target tier.}
\label{tab:implemented_features}
\footnotesize
\begin{tabular}{|l|c|c|}
\hline
\textbf{Feature} & \textbf{Evolution Status} & \textbf{Targeted Tier} \\ \hline
\href{https://scikgdash.orkg.org/overview}{\textit{\textcolor{link}{Predicates without semantic description}}} & MVP & Strategic \\ \hline
\href{https://scikgdash.orkg.org/overview}{\textit{\textcolor{link}{Classes without semantic description}}} & MVP & Strategic \\ \hline
\href{https://scikgdash.orkg.org/contributions}{\textit{\textcolor{link}{Contributions per entity tracker}}} & New Feature & Strategic \\ \hline
\href{https://scikgdash.orkg.org/visitors}{\textit{\textcolor{link}{Map of visitor navigation paths}}} & MVP & Analytical \\ \hline
\href{https://scikgdash.orkg.org/comments}{\textit{\textcolor{link}{Resource comment database}}} & MVP & Operational \\ \hline
\href{https://scikgdash.orkg.org/comparisons}{\textit{\textcolor{link}{Comparison metrics}}} & Extended & Operational \\ \hline
\href{https://scikgdash.orkg.org/predicates}{\textit{\textcolor{link}{Duplicate predicates pipeline}}} & Extended & Operational \\ \hline
\href{https://scikgdash.orkg.org/papers}{\textit{\textcolor{link}{Statements per paper}}} & MVP & Operational \\ \hline
\href{https://scikgdash.orkg.org/papers}{\textit{\textcolor{link}{Paper metadata analyzer}}} & New Feature & Operational \\ \hline
\end{tabular}%
\vspace{-0.2cm}
\end{table}

\noindent
\textbf{Strategic Tier -- Macro-Level Monitoring:} The \textit{Strategic Tier} focuses on graph-wide quality metrics to detect systemic compliance gaps and anomalies. This tier tracks \textit{predicates and classes without semantic descriptions}, allowing the C\&CB team to spot compliance issues at a glance. To ensure accountability and provenance in a crowdsourced environment, the newly introduced \textit{contributions per entity tracker} monitors the community's historical contribution frequencies across the entire ORKG to detect systemic irregularities, e.g., potentially low-quality bulk edits.\vspace{0.1cm}

\noindent
\textbf{Analytical Tier -- Proactive Usage Analytics:} The \textit{Analytical Tier} focuses on proactively deriving actionable design insights through analysis of human interaction with the ORKG, surfacing potential issues before they become apparent through individual contributions. The \textit{map of visitor navigation paths} uses anonymized usage data via the Matomo API to show user paths as interactive node-edge topologies. In this way, curators can analyze interaction patterns to identify structural dead ends or high bounce rates. The C\&CB team uses these behavioral insights to identify design issues in the ORKG and propose corresponding improvements to the development team. This feature transforms the dashboard from a reactive cleaning tool into a proactive design assistant.\vspace{0.1cm}

\noindent
\textbf{Operational Tier -- HITL Curation Workflows:} The \textit{Operational Tier} serves as the core interactive workspace where authenticated curators execute granular corrections using five workflows. To resolve predicate redundancy, the \textit{duplicate predicates pipeline} was extensively revised and expanded. It leverages semantic similarity matching and clustering algorithms over the PostgreSQL database to identify synonym or redundant predicates created due to non-restrictive crowdsourced modeling freedom. The user interface allows curators to safely merge or delete identified duplicate predicates from the ORKG production instance. Artifact-level completeness is addressed by \textit{statements per paper}, which measures statement density to identify under-modeled paper entries, and \textit{comparison metrics}, which crawls published matrix-based overviews to identify empty cells, wrongly assigned data types, or missing descriptions for targeted enrichment. Collaborative quality management is supported by the \textit{resource comment database}, allowing curators to attach administrative tags and review notes directly to RKG resources. Finally, bibliographic accuracy is supported by the \textit{paper metadata analyzer}, which enables curators to verify and update metadata by querying external registries using DOIs to make corrections.

\section{Interactive Demonstration Scenarios}
The demonstration of \textit{SciKGDash} at ISWC 2026 offers attendees the opportunity to try their hand at curation within the production environment. The demonstration is structured across the three functional tiers and reflects the work of the C\&CB team.\vspace{0.1cm}
\begin{enumerate}[noitemsep, leftmargin=12pt, topsep=0pt]
    \item \textbf{Scenario: Strategic Tier -- Macro-Level Monitoring.} Attendees monitor global quality metrics of the ORKG. They detect non-compliance issues by filtering for \textit{predicates and classes without semantic descriptions}. To detect systemic irregularities, they interact with the \textit{contributions per entity tracker}, monitoring community’s historical contribution frequencies. Attendees observe how sudden spikes in contribution density indicates a potential bulk import, demonstrating how macro-level monitoring helps to detect structural anomalies.
    \item \textbf{Scenario: Analytical Tier -- Proactive Usage Analytics.} Attendees focus on proactively deriving actionable design insight based on actual human interaction. They explore the \textit{map of visitor navigation paths}, which uses real-time, anonymized usage data via the Matomo API. The dashboard processes this raw data to render sequential user journeys as interactive chord diagram. Attendees track how users navigate through the ORKG to identify structural bottlenecks, dead ends, or high-bounce rates. They experience how these behavioral patterns are translated into change requests to improve the ORKG.
    \item \textbf{Scenario: Operational Tier -- HITL Curation Workflows.} Attendees make live corrections to the ORKG under supervision using the five operational workflows. First, they activate the \textit{duplicate predicates pipeline} to identify and resolve redundant properties flagged by semantic similarity clustering, executing a live merge or deletion. Next, they investigate artifact-level completeness by using \textit{statements per paper} to identify under-modeled entries and crawling published comparisons to compute \textit{comparison metrics}. Finally, attendees use the \textit{paper metadata analyzer} to verify and update metadata via automated DOI lookups. During all workflows, the attendees can attach administrative tags and review notes directly to RKG resources within the \textit{resource comment database}.
\end{enumerate}

\section{Evaluation Plan}
Building on the positive initial evaluation of the MVP~\cite{John_SciKGDash.2025}, we are currently planning a long-term, systematic evaluation of the new system. We organized the evaluation plan for \textit{SciKGDash} into short-, mid-, and long-term actions to balance immediate human-centric validation with large-scale verification.\vspace{0.1cm}

\noindent
\textbf{Short-term Actions:} The immediate phase focuses on assessing system adoption, user experience (UX), and cognitive load reduction among the members of the C\&CB team. Through structured usability tests, standardized UX assessments, and semi-structured interviews, we will examine how the curators rate the dashboard in terms of effectiveness, efficiency, and satisfaction.

\noindent
\textbf{Mid-term Actions:} The intermediate phase transitions to a quantitative assessment of the dashboards's backend analytical capabilities, focusing on the extensively reworked duplicate predicates pipeline. We will compute standard information retrieval metrics, including precision, recall, and F1-score, to evaluate the structural detection accuracy of our semantic similarity matching and clustering algorithms. These metrics will be benchmarked against a manually validated ground truth dataset established by the members of the C\&CB team. Additionally, tracking user-driven recommendation acceptance rates will provide a continuous feedback loop to systematically fine-tune automated clustering thresholds.

\noindent
\textbf{Long-term Actions:} The final phase comprises a longitudinal analysis of system logs and overall RKG quality during an extended period of use. By mining transactions and history logs, we will track long-term curator efforts, behavioral trajectories, and workflow adaptation patterns. The direct systemic impact of \textit{SciKGDash} on the ORKG ecosystem will be quantified by tracking macro-level data evolution trends over time, such as the distribution of chosen resolution strategies (e.g., frequency of hard-merges, soft-merges, or deletions) and the global improvement rates.

\section{Conclusion and Future Work}
In this paper, we presented \textit{SciKGDash}, a dashboard designed to scale HITL curation by the C\&CB team of the ORKG. The integration of strategic monitoring, analytical usage insights, and operational curation workflows bridges the gap between non-restrictive crowdsourced modeling and
semantic structural rigor essential for reliable, machine-actionable scientific knowledge. The dashboard has successfully matured from an MVP into a production-ready system, actively assisting curators in maintaining a high-quality knowledge graph.

As future work, our roadmap focuses on enhancing system autonomy, democratizing curation, and extending the RKG quality governance. Regarding system autonomy, we aim to fully leverage our PostgreSQL database as a persistent, primary data layer. This approach will decouple the whole dashboard from the ORKG's availability, ensuring data remains accessible during ORKG maintenance while simultaneously optimizing client-side performance by serving pre-computed, materialized views directly from the dashboard. Regarding democratizing curation, we will transition from an exclusive curator view to a hybrid model that enables users to interact with the dashboard in more detail. We plan to elevate users to become curators of their own contributions, providing them with personalized insights and performance metrics. By integrating leaderboard mechanisms and user-specific contribution tracking, we intend to gamify the curation process, thereby incentivizing the community to proactively maintain high RKG quality. Finally, we plan to extend our proactive design assistant by implementing automated, ad-hoc analytical workflows. Users will be able to input arbitrary URIs from the ORKG, e.g., a specific comparison URL, to trigger an incubation loop. The dashboard will compute a live quality score and generate automated, targeted recommendations for improvements, effectively pushing the RKG quality governance from reactive mechanisms to proactive guidance.


\begin{acknowledgments}
    This work was carried out as part of the SciD-QuESt subproject within the NFDIxCS consortium (project number 501930651), which is funded by the Deutsche Forschungsgemeinschaft (DFG).
\end{acknowledgments}

\section*{Declaration on Generative AI}
During the preparation of this work, the authors used Gemini 3.5 Flash to: Abstract drafting, Drafting content, Grammar and spelling check, Peer review simulation. After using this tool/service, the authors reviewed and edited the content as needed and take full responsibility for the publication’s content. 

\bibliography{references}




\end{document}